\documentclass[a4paper,pra,reprint, twocolumn,superscriptaddress]{revtex4}
\usepackage{ulem}
\usepackage{amssymb}
\usepackage{amsmath}
\usepackage{epsfig}
\usepackage{color}
\usepackage{graphics, graphicx}
\usepackage{bbold}
\usepackage{psfrag}
\usepackage{mathcomp}
\usepackage{amsmath}
\usepackage{amssymb}%
\usepackage{mathrsfs}%
\usepackage{subfigure}
\usepackage{verbatim}
\usepackage{xcolor}
\usepackage[colorlinks,citecolor=blue,urlcolor=blue]{hyperref}

\usepackage{mathrsfs}

\begin{document}

\date{\today}
\title{Probing two-body exceptional points in open dissipative systems}
\author{Peize Ding}
\affiliation{CAS Key Laboratory of Quantum Information, University of Science and Technology of China, Hefei 230026, China}
\author{Wei Yi}
\email{wyiz@ustc.edu.cn}
\affiliation{CAS Key Laboratory of Quantum Information, University of Science and Technology of China, Hefei 230026, China}
\affiliation{CAS Center For Excellence in Quantum Information and Quantum Physics, Hefei 230026, China}

\begin{abstract}
We study two-body non-Hermitian physics in the context of an open dissipative system depicted by the Lindblad master equation.
Adopting a minimal lattice model of a handful of interacting fermions with single-particle dissipation, we show that the non-Hermitian effective Hamiltonian of the master equation gives rise to two-body scattering states with state- and interaction-dependent parity-time transition. The resulting two-body exceptional points can be extracted from the trace-preserving density-matrix dynamics of the same dissipative system with three atoms. Our results not only demonstrate the interplay of PT symmetry and interaction on the exact few-body level, but also serve as a minimal illustration on how key features of non-Hermitian few-body physics can be probed in an open dissipative many-body system.
\end{abstract}

\maketitle

\section{Introduction}

Non-Hermitian physics has stimulated significant interest in recent years~\cite{PTreview1,PTreview2,PTreview3,Uedareview}, where particular attention has been devoted to its unconventional dynamics, peculiar critical behavior, and exotic band topology. A major driving force behind the booming field is the experimental implementation or simulation of these intriguing phenomena, particularly in open dissipative quantum systems~\cite{dalibard,daley,openbook}.  Therein, the system undergoes particle or energy loss to its environment, and a non-Hermitian effective Hamiltonian becomes relevant by imposing postselection~\cite{dalibard,daley}.
So far, a wide spectrum of non-Hermitian phenomena, ranging from parity-time (PT)-symmetry breaking and non-Hermitian criticality~\cite{benderreview,ptcrit,kznc}, to non-Hermitian skin effects and non-Bloch topology~\cite{WZ1,Budich,alvarez,mcdonald,ThomalePRB,Lee,WZ2,murakami}, have been experimentally implemented and explored in quantum mechanical systems such as the single-photon interferometry network~\cite{xuept,crit1,photonskin}, cold atoms~\cite{luole,bryce,yan,NHSOCexp,yanzeno}, nitrogen-vacancy centers~\cite{crit2,epencircle}, superconducting qubits~\cite{scpt}, and trapped ions~\cite{chenion,zhangion}. While most of these experiments investigate the single-particle aspects of the non-Hermitian physics, the interplay of non-Hermiticity and interaction is a fast-growing frontier with many open questions and fresh challenges~\cite{sf1,sf2,nonHtwobody,Yu,Cui}.

One of the key issues here is the relevance of non-Hermitian many-body Hamiltonians to open dissipative quantum systems~\cite{fu,michishita}.
While the latter is naturally characterized by trace-preserving density-matrix dynamics, the former requires a biorthogonal construction to ensure othornormality and recover the bosonic/fermionic statistics~\cite{DCB}.
Further, the postselection framework in a many-body setting requires an unchanged particle number~\cite{michishita}, thus exacting a stringent limit on the timescale within which the non-Hermitian description can be applied.
By contrast, this is not an issue with a non-interacting system, as is the case with the recent observations of PT transition and exceptional-point encircling in cold atoms~\cite{luole,NHSOCexp}. Therein,
a non-interacting atomic gas undergoes particle loss to the environment through optical pumping---a single-particle process. The dynamics is driven by a non-Hermitian effective Hamiltonian derived from the Lindblad master equation by dropping the quantum jump term $c\rho c^\dag$ (here $c$ is the atomic annihilation operator and $\rho$ the full density matrix of the non-interacting system).
This is equivalent to focusing only on atoms that are not lost to the environment, in the spirit of postselection.
Specifically, under the quantum-trajectory description~\cite{daley}, since dynamics of individual atoms are decoupled, they constitute an ensemble of independent trajectories with no quantum jumps, all driven by the non-Hermitian effective Hamiltonian.
Therefore, given a large number of atoms, the corresponding non-unitary dynamics can be probed by making measurements on the remaining atoms. Should interactions exist, however, dynamics of atoms in general would not decouple. Applying postselection would then amount to requiring the complete absence of quantum jumps for any single atom within the ensemble, which becomes exponentially unlikely with an increasing atom number.

Nevertheless, we demonstrate in this work that, key non-Hermitian physics can still be probed in the full density-matrix dynamics over fairly long times, provided the open many-body system be dominated by few-body correlations.
Using a minimal model of either two or three interacting fermions in a one-dimensional lattice, we first show that, under the non-Hermitian effective Hamiltonian of the corresponding Lindblad master equation, two-body scattering states of the system feature state- and interaction-dependent PT transitions. We then evolve the three-fermion dissipative system using the quantum trajectory scheme, taking into account the quantum jump processes. Remarkably, the decay of two-body correlations in this three-fermion open system follows the imaginary components of the complex eigenenergies of the two-body non-Hermitian scattering states. This enables us to extract the global exceptional point of the underlying two-body non-Hermitian system in the three-atom dissipative dynamics, from which the impact of interaction on the exceptional point is identified.
Our results suggest that, non-Hermitian many-body physic can in principle be probed in the context of an open dissipative many-body setting,
at least when both are dominated by few-body correlations.

Our paper is organized as follows. In Sec.~II, we present the model configuration, the corresponding Lindblad master equation, and the non-Hermitian effective Hamiltonian. We solve the two-body eigen problem of the non-Hermitian Hamiltonian using exact diagonalization in Sec.~III, where we reveal the existence of PT transitions and exceptional points in the scattering states. In Sec.~IV, we solve the Lindblad master equation for a three-fermion system using the quantum trajectory approach, and show the relevance between the two-body correlations therein and the complex eigenenergies of the non-Hermitian scattering states. We summarize in Sec.~V.

\section{Dissipative many-body system and non-Hermitian Hamiltonian}

\begin{figure}[tbp]
\includegraphics[width=0.4\textwidth]{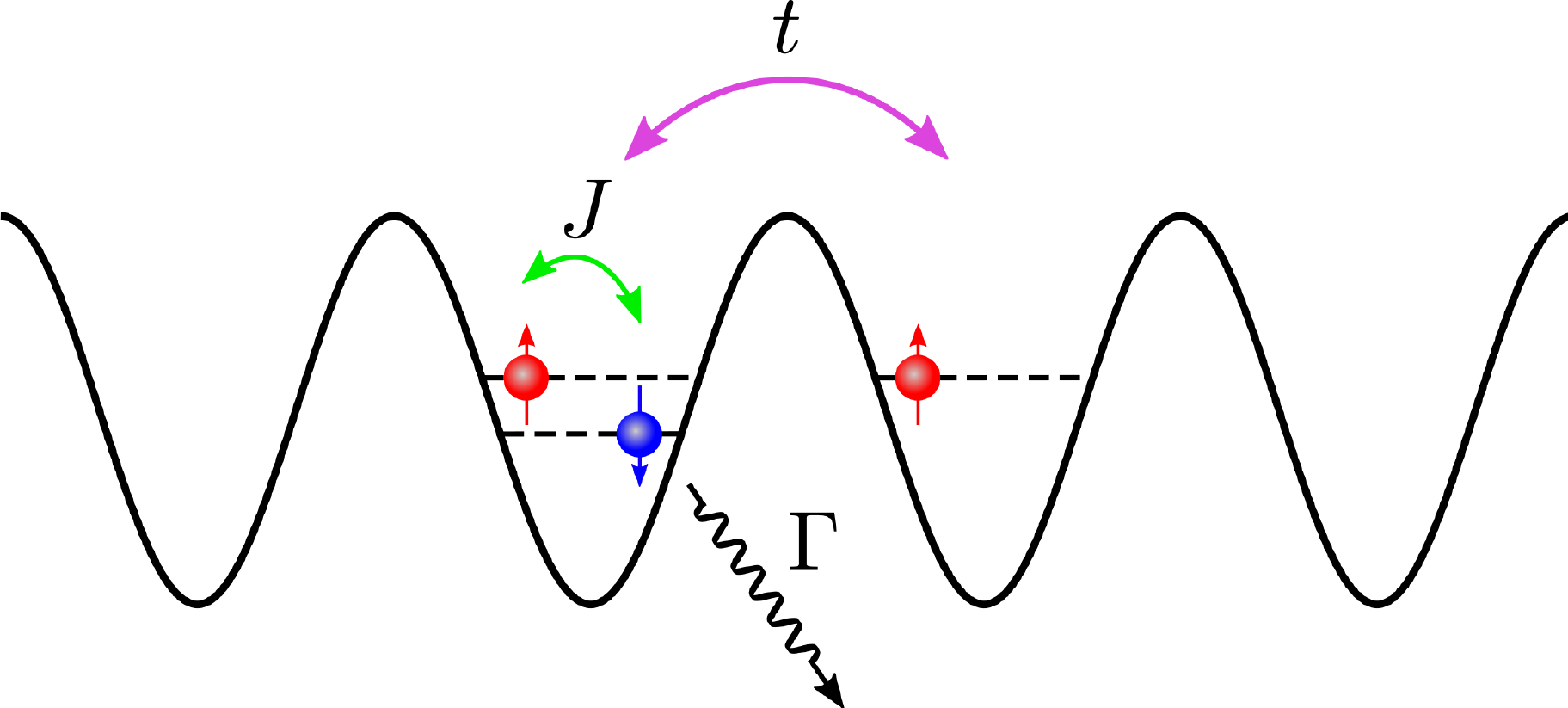}
\caption{Schematics of the dissipative lattice gas driven by the master equation (\ref{eq:lind}). The single-particle loss (with rate $\Gamma$) is induced by optical pumping, via an electronically excited state, to a third state (not drawn) that is not trapped by the lattice potential. The r.f. coupling rate $J$ and hopping rate $t$ are defined in the main text.
}
\label{fig:model}
\end{figure}

As illustrated in Fig.~\ref{fig:model}, we consider fermionic atoms with two hyperfine states in a one-dimensional optical lattice potential. The corrsponding Hamiltonian is given by
\begin{align}
H= & \sum_{k, \sigma} \epsilon_{k} c_{k, \sigma}^{\dagger} c_{k, \sigma}+J\sum_{k}\left(c_{k, \uparrow}^{\dagger} c_{k, \downarrow} + c_{k, \downarrow}^{\dagger} c_{k, \uparrow}\right)\nonumber \\
&+\frac{U_s}{\mathcal{L}} \sum_{k, k^{\prime}, q} c_{k+q, \uparrow}^{\dagger}
c_{k^{\prime}-q, \downarrow}^{\dagger} c_{k^{\prime}, \downarrow}
c_{k, \uparrow}\nonumber\\
&+\frac{U_p}{\mathcal{L}} \sum_{k, k^{\prime}, q, \sigma} \sin(\frac{k-k^\prime+2q}{2})\sin(\frac{k-k^\prime}{2})\nonumber\\
&\quad \quad\quad\quad\quad \times c_{k+q, \sigma}^{\dagger}
c_{k^{\prime}-q, \sigma}^{\dagger} c_{k^{\prime}, \sigma}
c_{k, \sigma}.
\label{eq:H}
\end{align}
Here $c_{k, \sigma} $ ($c^\dag_{k, \sigma} $) annihilates (creates) a fermionic atom with quasimomentum $k$ ($k\in [-\pi,\pi)$) in the hyperfine state $|\sigma\rangle$ ($\sigma=\uparrow,\downarrow$), $\epsilon_k=-t [\cos(ka_0)-1]$ with a hopping rate $t$ under the tight-binding approximation, $J$ is the radio-frequency (r.f.) coupling rate between different hyperfine spins, and $\mathcal{L}$ is the quantization length. We consider both $s$-wave and $p$-wave interactions, characterized by $U_s$ and $U_p$~\cite{pinteraction}, respectively, with $U_s,U_p<0$.

We further consider the case where one of the hyperfine spin states ($\left|\downarrow\right\rangle$) is subject to optical pumping, via an electronically excited state, out of the lattice potential. Under the Markovian approximation, the dynamics of the system is captured by the Lindblad master equation
\begin{align}
\frac{d \rho}{dt}=-i\Big(H_{\text{eff}}\rho-\rho H^\dag_{\text{eff}}\Big)+\Gamma\sum_{k}c_{k,\downarrow} \rho c_{k,\downarrow}^{\dagger},
\label{eq:lind}
\end{align}
where $\rho$ is the density matrix, the non-Hermitian effective Hamiltonian of the Lindblad equation is given by $H_{\text{eff}}=H-i\frac{\Gamma}{2}\sum_k c^\dag_{k,\downarrow}c_{k,\downarrow}$, and $\Gamma$ is the single-particle loss rate.

While the full dynamics of the open system is governed by Eq.~(\ref{eq:lind}), under the quantum trajectory framework, the dynamics is understood as a non-unitary time evolution driven by $H_{\text{eff}}$, which is further interrupted by quantum jumps $\{c_{k,\downarrow}\}$ with relative probabilities $\{\Gamma\delta t|c_{k,\downarrow}|\psi(t)\rangle|^2\}$. Here $\delta t$ is the coarse-grained time step, and $|\psi(t)\rangle$ is the instantaneous state of the system. It is often argued that, when the effects of quantum jumps are negligible, the open system would evolve under the non-Hermitian Hamiltonian $H_{\text{eff}}$. Such a postselection argument plays a key role in connecting realistic open quantum systems to the rich and exotic non-Hermitian physics that has attracted much attention of late~\cite{Uedareview}.

For a non-interacting atomic gas with $U_s=U_p=0$, imposing postselection is conveniently equivalent to focusing only on the dynamics of atoms that are not lost to the environment.
Since the full density matrix is just a direct product of single-particle density matrices, dynamics of each individual atom is driven by the same
master equation. Further, as atoms that remain necessarily have not undergone the quantum jump process, the trajectories of remaining atoms are
non-unitary evolutions driven by the same non-Hermitian effective Hamiltonian.
This is indeed the case with the recent experimental demonstrations of PT symmetry and exceptional-point encircling in cold atoms~\cite{luole,NHSOCexp}.

However, in an interacting system, the full density matrix can no longer be decomposed into single-particle ones.
Postselection thus corresponds to a complete absence of quantum jumps, i.e., it requires an unchanged total particle number.
The non-Hermitian effective Hamiltonian is then applicable at short times,
when the impact of quantum jump terms are negligibly small. This is equivalent to the practice in Ref.~\cite{Yu}, which projects the time evolution of the Lindblad equation onto the maximum-atom-number subspace. However, the corresponding time scale should become exponentially short with increasing particle number.
Generally, consider an $N$-particle system undergoing single-particle loss with the rate $\Gamma$.
The probability that not a single quantum jump occurs scales as $\sim e^{-N\Gamma \tau}$, where $\tau$ is the evolution time.
Therefore, the probability of all $N$ particles still remain at the time $1/\Gamma$ (for an evolution starting at $t=0$) is of the order $e^{-N}$, and the time scale at which the non-Hermitian effective Hamiltonian dominates should be $t< 1/N\Gamma$.
Nevertheless, we show in the following that, two-body physics under the non-Hermitian effective Hamiltonian {\it can} be probed in the full density-matrix dynamics of the corresponding Lindblad equation, on time scales that exceed $1/\Gamma$.

\section{Non-Hermitian two-body scattering state}

We first characterize the two-body problem under the non-Hermitian effective  Hamiltonian $H_{\text{eff}}$. To connect with previous cold-atom experiments on PT symmetry~\cite{luole,NHSOCexp}, we define the PT symmetric Hamiltonian $H_{\text{PT}}=H_{\text{eff}}+i\Gamma$. While the addition of the pure imaginary energy shift $i\Gamma$ does not change key physics such as the emergence and location of exceptional points, it renders Hamiltonian $H_{\text{PT}}$ PT symmetric in the non-interacting limit, with purely real (imaginary) eigenspectrum for $J>\Gamma/4$ ($J<\Gamma/4$).

\begin{figure}[tbp]
	\centering
	\includegraphics[width=0.45\textwidth] {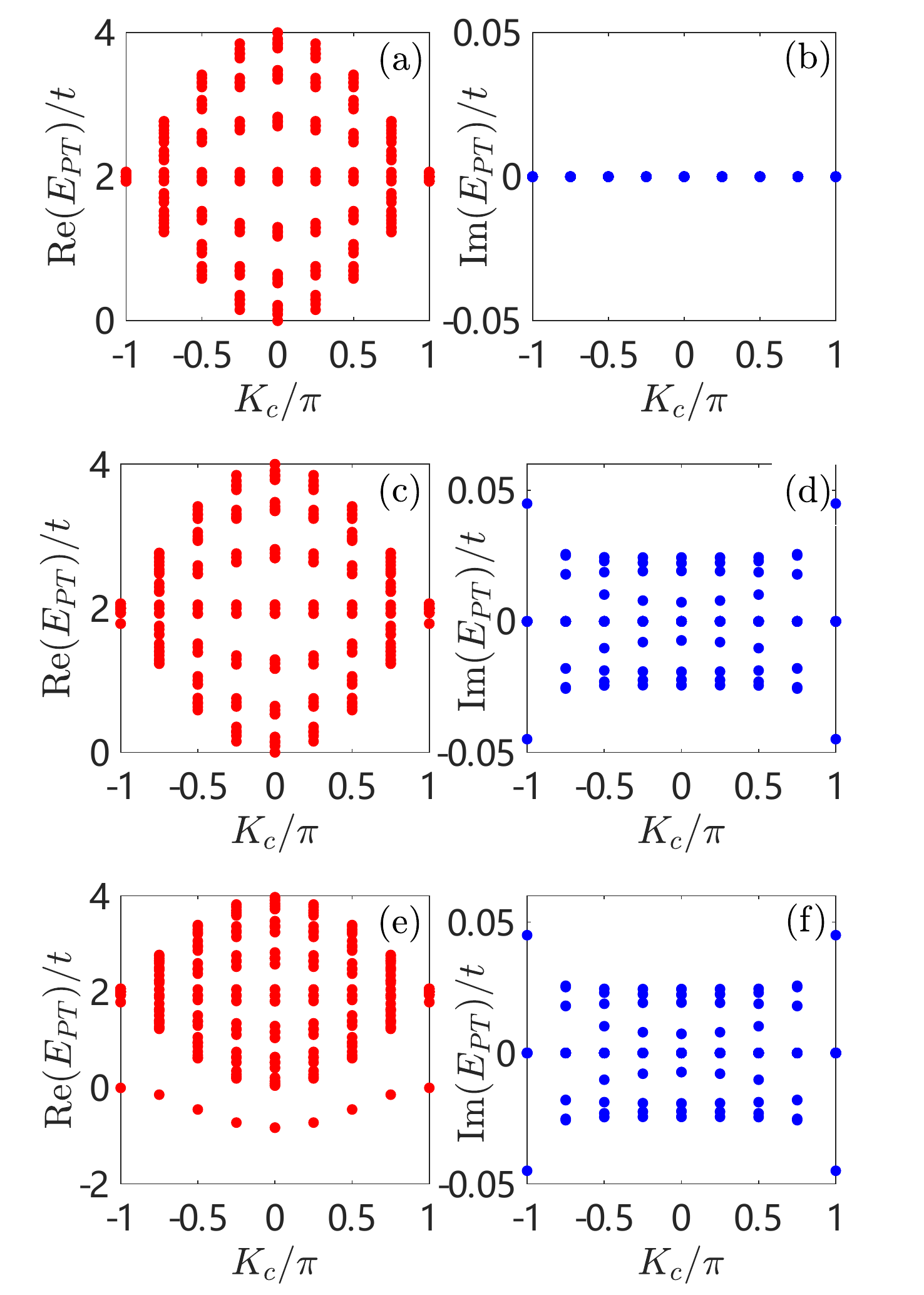}
	\caption{(Color online) Complex energy spectra of two fermions under the PT symmetric Hamiltonian $H_{\text{PT}}$ for (a) $ U_s=0$, $U_p=0$, $J/t=0.04 $; (b) $ U_s=0$, $U_p/t=-0.2$, $J/t=0.04$; and (c) $ U_s/t=-2$, $U_p/t=-0.2$, $J/t=0.04 $. The left (right) panel shows the real (imaginary) components of the eigenspectra, with $K_c$ the center-of-mass momentum of the two-body state. For all numerical calculates, we take a one-dimensional lattice with $N=16$ sites.}
\label{fig:fig1}
\end{figure}

In Fig.~\ref{fig:fig1}, we show the numerically evaluated eigenspectra for two fermions along a lattice of $N=16$ sites, with the parameters $J/t=0.04$ and $\Gamma/t=0.1$. Here $K_c$ is the center of mass of the two-body state, which is a good quantum number of the system. In the non-interacting case [see Fig.~\ref{fig:fig1}(a)(b)], $H_{\text{PT}}$ is in the PT-unbroken regime, with purely real eigenspectra. The PT symmetry is broken under a sufficiently large $p$-wave interaction, as the eigenspectra acquire imaginary components under a finite $U_p$ [see Fig.~\ref{fig:fig1}(c)(d)]. This in contrast to the $s$-wave interaction, which does not affect the imaginary components of the eigenspectra [see Fig.~\ref{fig:fig1}(e)(f)]. Note that while discrete two-body bound states can be identified, for instance in Fig.~\ref{fig:fig1}(e), it is the two-body scattering states within the continuum that acquire imaginary components.
Importantly, we expect the PT transition point (or the exceptional points) of the non-interacting Hamiltonian be shifted by the $p$-wave interaction.

\begin{figure}[tbp]
	\centering
	\includegraphics[width=0.45\textwidth] {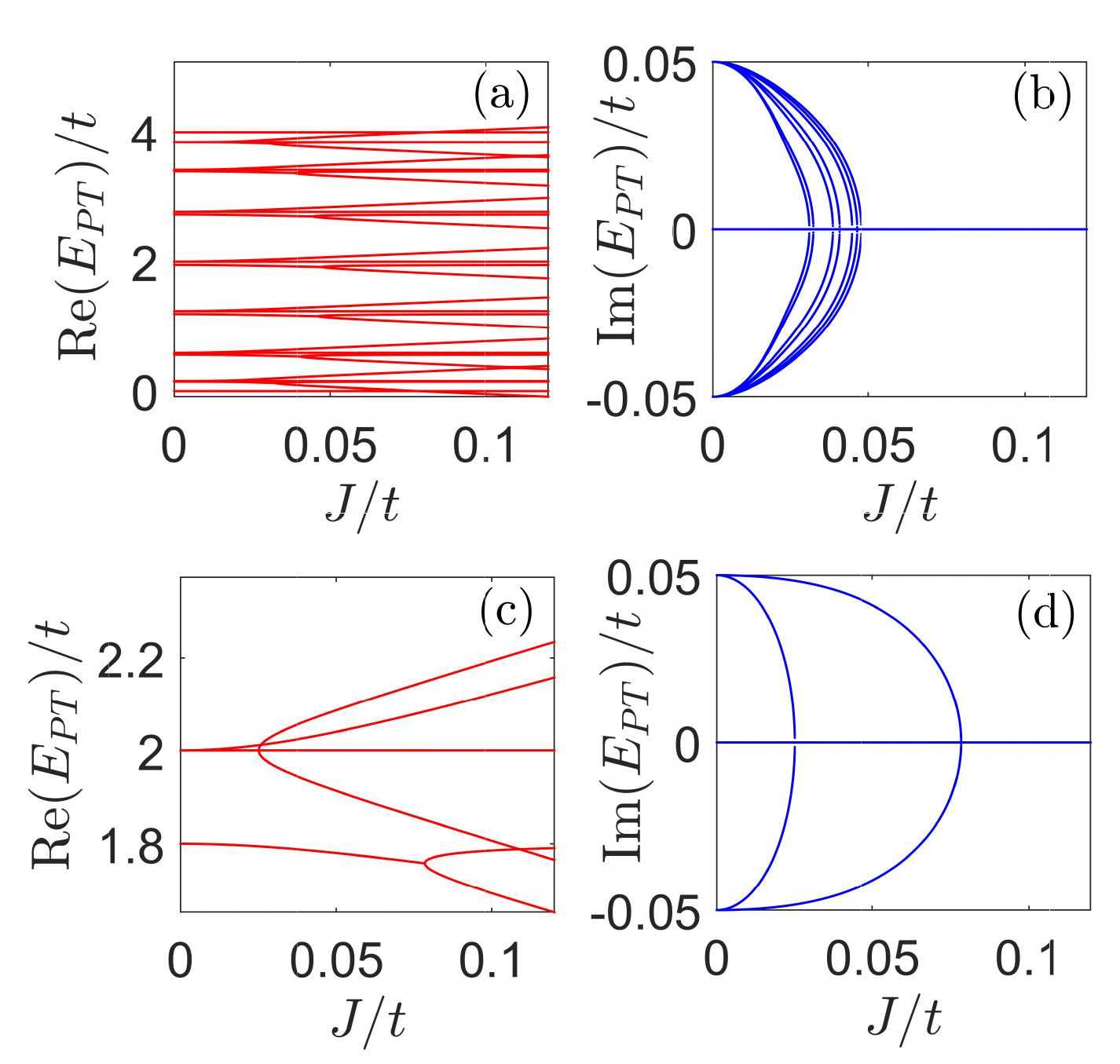}
	\caption{(Color online) Complex energy spectra of $ H_{\text{PT}} $ as functions of $ J/t$ for (a) $ K_c=0$ and (b) $ K_c=\pi $. As $ J/t$ decreases, the scattering states coalesce in pairs at state-dependent exceptional points, through which their real components merge and imaginary components bifurcate. The full eigenspectra become completely real above a critical $J_c/t$, which is identified as the global exceptional point (or the global PT transition point).
}
\label{fig:fig2}
\end{figure}

\begin{figure}[tbp]
	\centering
	\includegraphics[width=0.45\textwidth] {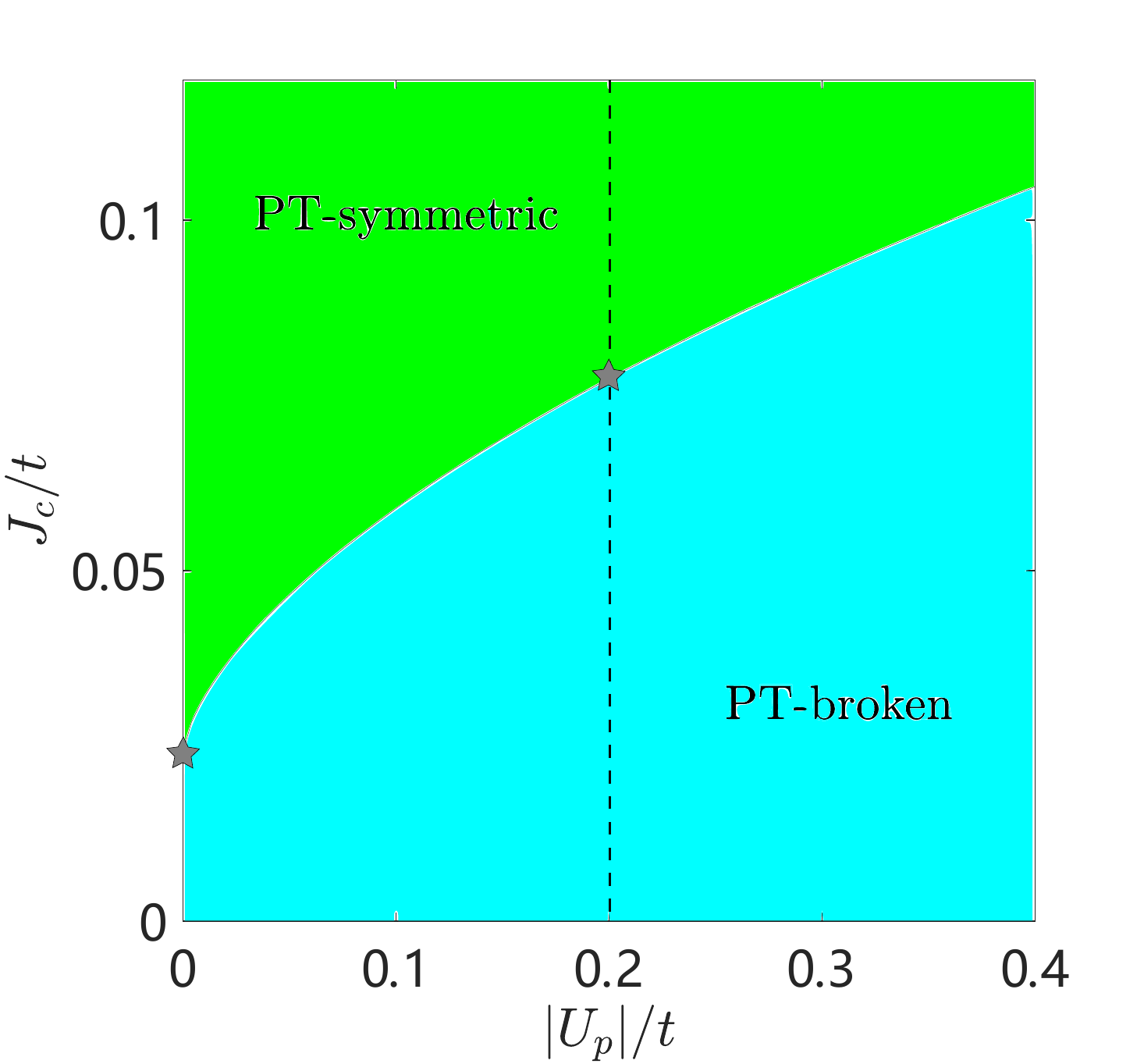}
	\caption{(Color online) PT phase diagram for $H_{\text{PT}}$ with $U_s=0$ and $U_p<0$. The global exceptional point $J_c/t$ increases with larger $p$-wave interaction.
The parameter used in Fig.~\ref{fig:fig2} is indicated by the vertical dashed line.}
\label{fig:fig3}
\end{figure}

This is confirmed in Figs.~\ref{fig:fig2} and \ref{fig:fig3}. Specifically, in Fig.~\ref{fig:fig2}, we show the splitting of exceptional points in different $K_c$ sectors under a finite $U_p$. With increasing $J$, scattering states sequentially coalesce in pairs at an array of second-order exceptional points. We identify the exceptional point with the largest $J$ as the global PT transition point under the $p$-wave interaction,denoted by $J_c$. The resulting PT phase diagram is shown in Fig.~\ref{fig:fig3}. Apparently, $p$-wave interactions shift the global exceptional point toward larger $J$, consistent with the results in Fig.~\ref{fig:fig1}.

\begin{figure}[tbp]
	\centering
\includegraphics[width=0.45\textwidth] {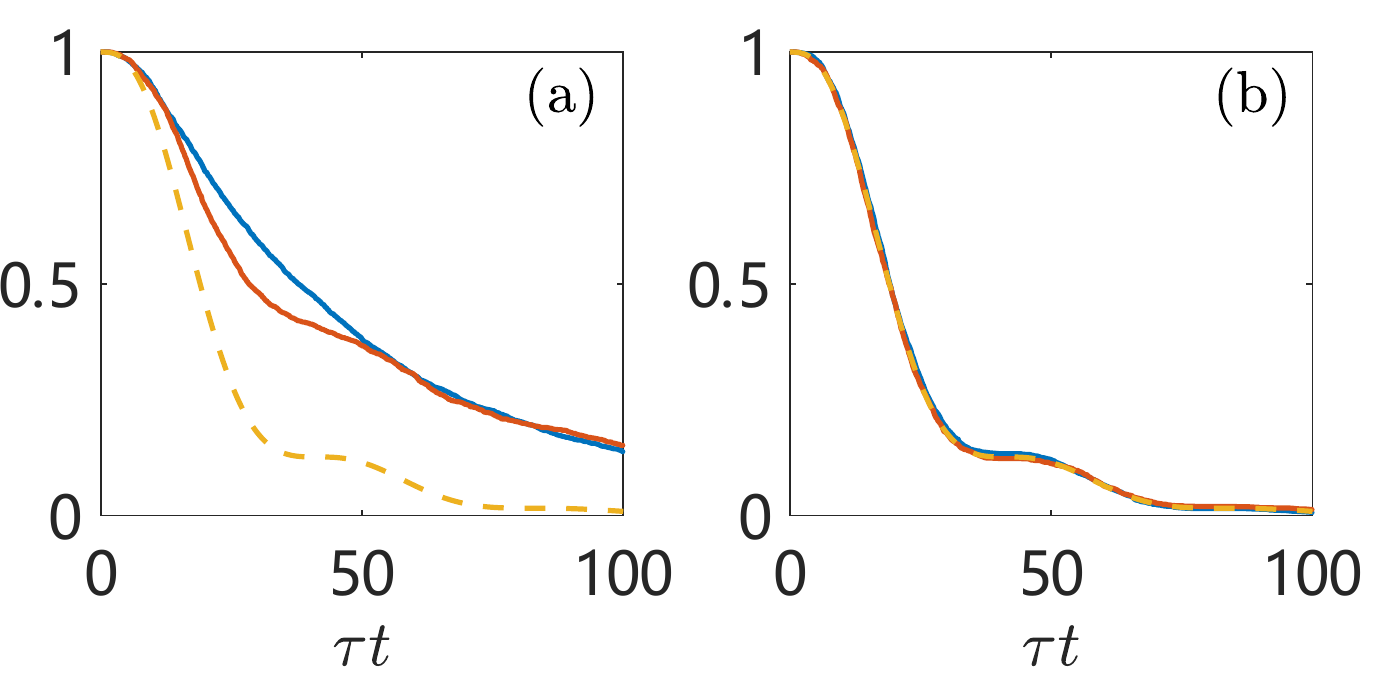}
\caption{(Color online) Comparison of the total particle-number evolution $\sum_{k,\sigma}\text{Tr}\Big[ \rho(\tau)c^\dag_{k,\sigma}c_{k,\sigma}\Big]$ under the Lindblad master equation for systems initialized with
three atoms (blue solid), two atoms (red solid), and a single atom (yellow dashed), respectively, in the states $|\uparrow,k=-\pi\rangle\otimes |\uparrow,k=0\rangle\otimes |\uparrow, k=\pi-2\pi/N$ ($N=8$), $|\uparrow,k=\pi\rangle\otimes |\uparrow,k=0\rangle$, and $|\uparrow,k=0\rangle$. The interaction parameters are (a) $U_p/t=-0.5$, (b) $U_p=0$, with $U_s=0$, $J/t=0.08$, and $\gamma/t=0.1$ for both cases. To facilitate comparison, the particle numbers in all cases are normalized to one.
}
\label{fig:N}
\end{figure}

\section{Probing non-Hermitian physics in open system}

We now show that the global exceptional point $J_c$ of the two-body scattering states in Fig.~\ref{fig:fig3} can be probed from the density-matrix dynamics under the full Lindblad equation. Under interactions, information of the two-body exceptional point is difficult to extract from the particle-number dynamics of a many-body system. This is illustrated in Fig.~\ref{fig:N}, where we compare the particle-number evolution under the Lindblad master equation for systems initialized with different particle numbers, either with (Fig.~\ref{fig:N}(a)) or without (Fig.~\ref{fig:N}(b)) interactions.
While the dynamics for different initial particle numbers appear to be the same without interactions, they generally differ under a finite $U_p$.
This shows that non-Hermitian two-body physics cannot be directly probed using particle-number dynamics in a three-body dissipative system. Note that non-Hermitian two-body physics can be fully captured by a Lindblad equation initialized in the two-body sector, since quantum jump terms in a two-body sector only couple to three-body states~\cite{Yu}.

Instead, other observables should be adopted. For a minimal demonstration, we solve the Lindblad equation in a three-fermion system, initialized in the state $|\psi^{(S)}(K_c=-\pi)\rangle\otimes |\uparrow,k=\pi-2\pi/N\rangle$. Here $|\uparrow,k\rangle$ is a single-particle state with hyperfine spin $|\uparrow\rangle$ and momentum $k$, $|\psi^{(S)}(K_c)\rangle$ is the two-body scattering state of $H_{\text{eff}}$ with a center-of-mass momentum $K_c$, and $|\psi^{(S)}(K_c=\pi)\rangle$ is the eigenstate with the largest imaginary eigenenergy component, denoted as $\text{Im}(E^{(S)})$. We take the system size $N=8$ for numerical calculations.
Further, we define a normalized two-body correlation function
\begin{align}
G_{\alpha\beta}(k_1,k_2)=\frac{\text{Tr} \Big[\rho(\tau) c_{k_1,\alpha}^{\dagger} c_{k_2,\beta}^{\dagger} c_{k_2,\beta} c_{k_1,\alpha}\Big]}{\text{Tr} \Big[\rho(0) c_{k_1,\alpha}^{\dagger} c_{k_2,\beta}^{\dagger} c_{k_2,\beta} c_{k_1,\alpha}\Big]}. \label{eq:G}
\end{align}

\begin{figure}[tbp]
    \centering
   \includegraphics[width=0.45\textwidth] {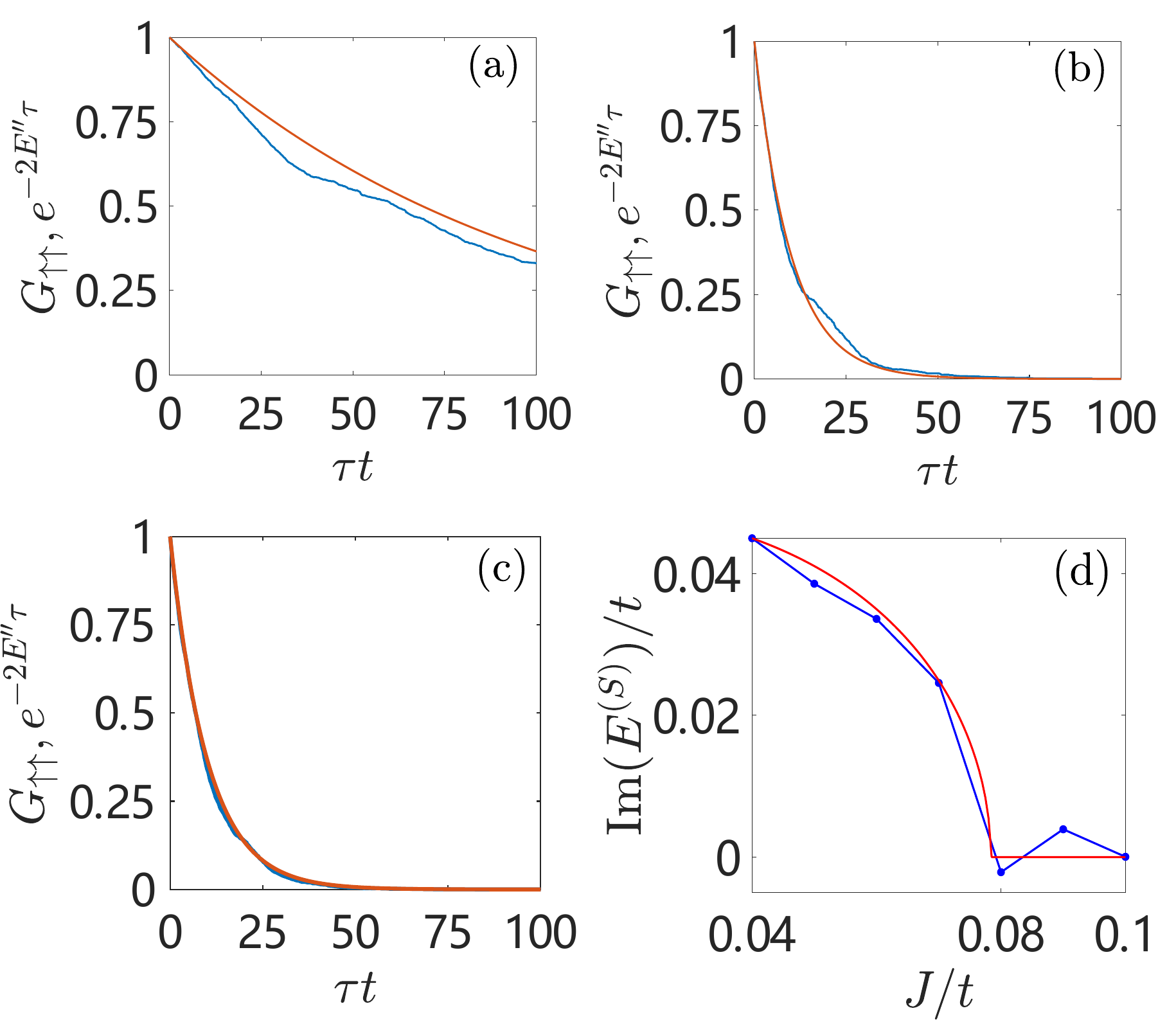}
   \caption{(Color online) (a)(b)(c) Time evolution of the normalized two-body correlation function $ G_{\uparrow\uparrow}(k_1=-\pi,k_2=0)$ (blue lines) under the quantum trajectory approach for (a) $ J/t=0.04 $, (b) $ J/t=0.08 $, and (c) $ J/t=0.1 $. We take $ U_p/t=-0.2$, $U_s=0$, and $\Gamma/t=0.1 $ for numerical calculations.
   The red lines show the time evolution of the norm of the corresponding two-body eigenstate of the non-Hermitian Hamiltonian $ H_{\text{PT}} $.
   (d) Comparison between the imaginary component of the two-body eigenenergy under $ H_{\text{PT}} $ (red line), with the numerically fitted exponent of the correlation-function decay (blue line and symbol). For the exponent, we numerically fit the time-dependent correlation function up to the time $\tau t=5$. For all our numerical calculations here, we take the lattice size $N=8$. We average over $2000$ trajectories for the quantum-trajectory calculations.}
      \label{fig:fig4}
\end{figure}

We evolve the Lindblad equation using the quantum trajectory approach, and plot the correlation function $ G_{\uparrow\uparrow}(k_1=-\pi,k_2=0) $ in Fig.~\ref{fig:fig4}.
Due to the particular choice of initial state and the two-body correlation function, the decay of the correlation function follows the imaginary component of the two-body scattering state with the largest critical $J$.
Thus, by fitting the exponents of decay in $G$, we are able to map out the global exceptional points in the phase diagram Fig.~\ref{fig:fig3}, and retrieve key properties of a two-body non-Hermitian Hamiltonian from the full dynamics of a interacting three-body open system.
We expect that this would hold true for many-body open systems, as long as the dominant correlations remain few-body in nature.

Finally, we note that, should we choose a different initial state and two-body correlation function, we would be able to extract information of other two-body scattering states.

\section{Summary and Discussion}\label{sec:sum}

Adopting a minimal model of a few dissipative fermions on a one-dimensional lattice, we show that PT transitions exist in the scattering states of the non-Hermitian Hamiltonian, and are shifted by the $p$-wave inter-atomic interactions. The interaction-shifted global exceptional point can be probed by measuring two-body correlations in the trace-preserving density-matrix dynamics driven by the Lindblad master equation. We therefore explicitly demonstrate a minimal scenario where key properties of a non-Hermitian interacting Hamiltonian can be probed in the context of an open system.

In particular, while one expects the non-Hermitian many-body Hamiltonian to be of relevance on a short time scale of $1/N\Gamma$ (see discussions in Sec.~II), it is remarkable that the two-body correlation captures key features of the two-body non-Hermitian scattering states at time scales even longer than $1/\Gamma$ (see Fig.~\ref{fig:fig4}). We attribute such a phenomenon to the dominance of few-body correlations in the open dissipative system. Our result is complementary to previous attempts at connecting non-Hermitian many-body Hamiltonians with open dissipative systems~\cite{fu,michishita}, and is relevant to cold atomic gases where few-body correlations dominate.

\section*{Acknowledgements}
We thank Xiaoling Cui for helpful discussions.
This work has been supported by the Natural Science Foundation of China (Grant No. 11974331) and the National Key R\&D Program (Grant Nos. 2016YFA0301700, 2017YFA0304100).

\end{document}